\documentclass[structabstract, longauth, regular]{aa} 
\renewcommand{\epsilon}{\varepsilon}

\usepackage{graphicx}
\usepackage{verbatim}
\usepackage{txfonts}
\usepackage[switch]{lineno}
\usepackage{setspace}
\usepackage{natbib}
\bibpunct{(}{)}{;}{a}{}{,} % to follow the A&A style

\usepackage{color}

\newcommand{\lsi}{LS~I~+61$^{\circ}$303}

\begin{document}

\title{Super-orbital variability of \lsi\ at TeV energies}
%\title{Multi-year studies of the gamma-ray binary \lsi\ performed with the MAGIC telescopes }
%\title{Multi-year study and super-orbital modulation signatures in the TeV flux of the gamma-ray binary \lsi\ observed with the MAGIC telescopes }

\author{
M.~L.~Ahnen\inst{1} \and
S.~Ansoldi\inst{2} \and
L.~A.~Antonelli\inst{3} \and
P.~Antoranz\inst{4} \and
A.~Babic\inst{5} \and
B.~Banerjee\inst{6} \and
P.~Bangale\inst{7} \and
U.~Barres de Almeida\inst{7,}\inst{25} \and
J.~A.~Barrio\inst{8} \and
J.~Becerra Gonz\'alez\inst{9,}\inst{26}
% ...
}
\institute { ETH Zurich, CH-8093 Zurich, Switzerland
\and Universit\`a di Udine, and INFN Trieste, I-33100 Udine, Italy
\and INAF National Institute for Astrophysics, I-00136 Rome, Italy
\and Universit\`a  di Siena, and INFN Pisa, I-53100 Siena, Italy
\and Croatian MAGIC Consortium, Rudjer Boskovic Institute, University of Rijeka, University of Split and University of Zagreb, Croatia
\and Saha Institute of Nuclear Physics, 1/AF Bidhannagar, Salt Lake, Sector-1, Kolkata 700064, India
\and Max-Planck-Institut f\"ur Physik, D-80805 M\"unchen, Germany
\and Universidad Complutense, E-28040 Madrid, Spain
\and Inst. de Astrof\'isica de Canarias, E-38200 La Laguna, Tenerife, Spain; Universidad de La Laguna, Dpto. Astrof\'isica, E-38206 La Laguna, Tenerife, Spain
\and University of \L\'od\'z, PL-90236 Lodz, Poland
\and Deutsches Elektronen-Synchrotron (DESY), D-15738 Zeuthen, Germany
\and Institut de Fisica d'Altes Energies, The Barcelona Institute of Science and Technology (IFAE-BIST), Campus UAB, 08193 Bellaterra (Barcelona), Spain
\and Universit\"at W\"urzburg, D-97074 W\"urzburg, Germany
\and Universit\`a di Padova and INFN, I-35131 Padova, Italy
\and Institute for Space Sciences (CSIC/IEEC), E-08193 Barcelona, Spain
\and Centro de Investigaciones Energ\'eticas, Medioambientales y Tecnol\'ogicas, E-28040 Madrid, Spain
\and Technische Universit\"at Dortmund, D-44221 Dortmund, Germany
\and Unitat de F\'isica de les Radiacions, Departament de F\'isica, and CERES-IEEC, Universitat Aut\`onoma de Barcelona, E-08193 Bellaterra, Spain
\and Universitat de Barcelona, ICC, IEEC-UB, E-08028 Barcelona, Spain
\and Japanese MAGIC Consortium, ICRR, The University of Tokyo, Department of Physics and Hakubi Center, Kyoto University, Tokai University, The University of Tokushima, KEK, Japan
\and Finnish MAGIC Consortium, Tuorla Observatory, University of Turku and Department of Physics, University of Oulu, Finland
\and Inst. for Nucl. Research and Nucl. Energy, BG-1784 Sofia, Bulgaria
\and Universit\`a di Pisa, and INFN Pisa, I-56126 Pisa, Italy
\and ICREA and Institute for Space Sciences (CSIC/IEEC), E-08193 Barcelona, Spain
\and now at Centro Brasileiro de Pesquisas F\'isicas (CBPF/MCTI), R. Dr. Xavier Sigaud, 150 - Urca, Rio de Janeiro - RJ, 22290-180, Brazil
\and now at NASA Goddard Space Flight Center, Greenbelt, MD 20771, USA and Department of Physics and Department of Astronomy, University of Maryland, College Park, MD 20742, USA
% ...
}

\date{Received ... , 2015; accepted ... , 2015}

%%%%%%%%%%%%%%%%%%%%%%%%%%%%%%%%%%%%%%%%%
\abstract
{The gamma-ray binary \lsi\ is a well-established source from centimeter radio up to very high energy (VHE; E $>$ 100 GeV). The broadband emission shows a periodicity of $\sim$26.5 days, coincident with the orbital period. A longer (super-orbital) period of 1667 $\pm$ 8 days was proposed from radio variability and confirmed using optical and high-energy (HE;  E > 100 MeV) gamma-ray observations. In this paper, we report on a four-year campaign performed by MAGIC together with archival data concentrating on a search for a long-timescale signature in the VHE emission from \lsi.}
{ We  focus on the {search for super-orbital} modulation of the VHE emission, similar to that observed at other energies, and on the search for correlations between TeV emission and an optical determination of the extension of the circumstellar disk.
%This paper has tree aims. First, the search for super-orbital modulation in the VHE flux regime of this source, as found in other wavelengths. Second, the search for correlation with the TeV emission and optical data taken simultaneously. Finally, the pursuit of spectral features for different conditions in the binary system. 
 }
{A four-year campaign has been carried out {using the MAGIC telescopes}. The source was observed during the orbital phases when the periodic VHE outbursts have occurred ($\phi$ = 0.55 -- 0.75, {one orbit = 26.496 days}). Additionally, we
 included archival MAGIC observations and data published by the VERITAS
collaboration in these studies. For the correlation studies, \lsi\ has
also been observed during the orbital phases where sporadic VHE emission had
been detected in the past ($\phi$ = 0.75 -- 1.0). These MAGIC observations were
simultaneous with optical spectroscopy from the LIVERPOOL telescope. 
}
{The TeV flux of the periodical outburst in orbital phases $\phi$ = 0.5 -- 0.75 was found to {show yearly variability} consistent with the long-term modulation of $\sim$4.5 years found in the radio band. This modulation of the TeV flux can be well
described by a {sine function} with a best-fit period of $1610\pm 58$ days. The complete data, including archival observations, span two super-orbital periods. There is no evidence for a correlation between the TeV emission and the mass-loss rate of the Be star, but this may be affected by the strong, short-timescale (as short as intra-day) variation displayed by the H$\alpha$ fluxes.  

}
{}
%%%%%%%%%%%%%%%%%%%%%%%%%%%%%%%%%%%%%%%%%


\titlerunning{Multi-year study of \lsi}
\maketitle

%%%%%%%%%%%%%%%%%%%%%%%%%%%%%%%%%%%%%%%%%
\section{Introduction}
%%%%%%%%%%%%%%%%%%%%%%%%%%%%%%%%%%%%%%%%%
%\linenumbers
%\doublespacing

\lsi\ {(= V615 Cas )} is a gamma-ray binary composed of a rapidly
rotating Be star of spectral type B0Ve \citep{Hutchings81} with a circumstellar
disk and a compact object of unknown nature. The compact object, either a neutron star
(NS) or a stellar-mass black hole (BH), has an eccentric orbit ($e = 0.54 \pm 0.03$) with a period of
26.4960(28) days, determined from radio observations \citep{Gregory02}. The periastron
passage occurs at phase $\phi=0.23 \pm 0.03$, although its precise timing 
depends on the orbital solution \citep{Gregory02, Casares2005, Grundstrom07, Aragona09}.\footnote{Here, the reference for the periastron passage is taken at $T_{0}$ = 43366.275 MJD \citep{Gregory02}.} 

%at a distance of 2.0 $\pm$ 0.2 kpc \citep{Frail},
%[NB: there's no reason to discuss the distance here if it's not used, reference it for luminosity when needed later

\lsi\ is one of the few binary systems detected from radio to
{VHE gamma rays}. The origin of the broadband emission
 is still under debate. \lsi\ was proposed as a microquasar, based on its extended jet-like radio-emitting structures, see \citet{Massi_radio}, for instance, and more recently, \cite{Massi2014}. {However}, images obtained with the VLBA during a
complete orbital cycle showed a rotating tail-like morphology of
overall size 5 -- 10 mas \citep{Dhawan06}, consistent with a 
pulsar wind scenario \citep{Maraschi81}.  Similar phase-resolved structures were later observed by \cite{Albert08_LSI}. 
The source appears extended in X-rays \citep{Paredes07, Kargaltsev2014}.
{Deep observations searching} for X-ray pulsing have yielded only upper limits \citep{ReaPul}.

\lsi\ was first detected in the VHE regime by MAGIC in 2006
\citep{Albert_Science_LSI}. {During the following years, more} {} observations have been performed by the
MAGIC \citep{Albert08_LSI, Anderhub09, MAGIC_lsi_periodic09, lsi_flux_states} and VERITAS collaborations \citep{VERITAS08}. {The VHE emission region is unresolved ($< 0.1 \deg$)}. The system is also a HE gamma-ray \textit{Fermi}-LAT source \citep{Fermi_lsi}. Long-term monitoring shows that the HE emission undergoes periodic outbursts slightly after the periastron passage, around phases $\phi$ $\sim$
0.3 -- 0.45 \citep{Hadasch12}.  
%The system is also a {high energy (HE;  > 100 MeV )} gamma-ray \textit{Fermi}-LAT source \citep{Fermi_lsi}. Long-term monitoring shows that the HE emission undergoes periodic outbursts slightly after the periastron passage, around phases $\phi$ $\sim$0.3 -- 0.45 \citep{Hadasch12}.  
{In the orbit from MJD 53752.7 until MJD 53779.2} (January - February 2006), a TeV peak was first detected at phases
$\phi$ $\sim$ 0.6 -- 0.7 (which correspond to the phases next to the apastron)
at a level of $\sim$16\% of the Crab Nebula flux for energies above 400 GeV. However, in October 2009 -- January 2010, the source was in a low flux state \citep{lsi_flux_states} and the emission peak was above 300 GeV with a flux of only 5.4$\%$ of the
Crab Nebula.  Despite the state of the system, the highest TeV flux also occurred at phases $\phi$ = 0.6 --
0.7, consistent with previous observations. The spectral fit parameters agreed with our previous determinations \citep{Albert08_LSI, Anderhub09, MAGIC_lsi_periodic09}. 
Significant emission has also occasionally been observed at phases $\phi$ = 0.8 --
1, where MAGIC detected the system at $\sim$4\% Crab Units (C.U). \citep{MAGIC_lsi_periodic09}.  Emission around periastron has been reported once, by VERITAS, in late 2010 at phase $\phi$ = 0.081 \citep{Acciari11} when the compact star was {near superior conjunction}.

Orbit-to-orbit variability has been associated with a super-orbital modulation. This was first proposed based on centimeter radio variations that show approximately sinusoidal modulation over 1667 $\pm$ 8 days \citep{Paredes1987, Gregory02}. A similar long-term behavior has recently been suggested {for X-rays} {(3 -- 30\,keV observed with \textit{RXTE};} \cite{Li_12,Chernya2012}), hard X-rays {(18 -- 60\,keV observed with \textit{INTEGRAL};} \cite{Li_14}) 
and HE gamma-rays {(100 MeV -- 300\,GeV observed with Fermi/\textit{LAT};} \cite{Fermi2013}). 
In the LAT energy range, the super-orbital variability is almost invisible around the periastron, where the compact object is inside (or highly affected by) the Be circumstellar disk, but appears around apastron.

Two short ($<0.1$\,s) and very luminous ($>10^{37}$\, erg\,s$^{-1}$) bursts have been detected from the direction of \lsi\ by the \textit{Swift} Burst Alert Telescope \citep{Barthelmy08,Burrows12}. A detailed analysis of these events was presented by \citet{Torres12}, where their similarities to those observed from magnetars were analyzed. 
If the bursts are {associated with the system, as it seems, it requires} that the system harbors a neutron star and not a black hole.

The idea of a transitioning system, that is, a system that swings from one state to another at every orbital period, 
has been suggested by \cite{Zamanov2001}. 
{\cite{Torres12} and \cite{Papitto_2012}} extended this to the super-orbital behavior and 
analyzed orbitally induced variability. 

In this so-called flip-flop scenario, the system changes from propeller
regime accretion at periastron, where the pulsar magnetosphere is
disrupted, to an ejector regime (rotational-powered pulsar) at apastron, when
particles are accelerated to TeV energies in the inter-wind shock {formed
at the collision region between the neutron star and the stellar wind.}
Despite the system
differences, this would make \lsi\ similar to the transitional pulsars in
redbacks, for example, \citet{Archi2009} and \citet{Papitto2013}.

{Changing the Be star mass-loss rate can cause the switching from one state to the other to vary in orbital phase, modulated along the super-orbital timescale as measured by the equivalent width (EW) of the H$\alpha$ line, for instance. The sizes of
the stellar disks of Be stars correlate with the EW of the
H$\alpha$ emission line (e.g., see \citet{Porter2003} and \citet{Reig2011} for a
review). \cite{Grundstrom_Gies2006} also showed that the EW  is correlated with the radius of the circumbinary disk, and therefore it can be used as a proxy of the latter.
For times of high mass-loss rate, at which the
influence of the disk matter upon the compact object can be felt along a greater portion 
 of the orbit, the propeller regime can be operative even close to
apastron. If this is the case, the TeV emission of \lsi\ would be reduced \citep{Torres12}.}

%[NB: You don't need the next lines, just start the paper.]
%In this paper we present the long-term behavior of \lsi, by looking for flux and spectral variability at timescales of years.
%In particular, we search for super-orbital modulation of the VHE flux and put our results in the context of the predictions of the flip-flop model.

%%%%%%%%%%%%%%%%%%%%%%%%%%%%%%%%%%%%%%%%%
\section{Observations}
%%%%%%%%%%%%%%%%%%%%%%%%%%%%%%%%%%%%%%%%%

A campaign {using the MAGIC telescopes} together with the  optical telescope was carried
out over four years.
Both instruments are located on
the island of La Palma, in the Canary Islands, Spain, at the observatory of El
Roque de Los Muchachos (28$^\circ$N, 18$^\circ$W, 2200 m above the sea level).
The MAGIC telescopes are two Imaging Atmospheric Cherenkov
Telescopes (IACTs) with diameters of 17\,m, each one with a pixelized camera, covering a field of view of $\sim3.5^{\circ}$. The current
sensitivity of the stereoscopic MAGIC telescopes is 0.71$\%$ $\pm$ 0.02$\%$ of
the Crab Nebula flux in 50 h of observation for energies above 250 GeV
\citep{MAGIC_performance_2014}. The angular resolution at these energies is
$\lesssim 0.1^\circ$ (68$\%$ containment radius) and the energy resolution is $\sim18$\%. For
monoscopic observations (also referred to as mono-observations) the integral
sensitivity above 280 GeV is about 1.6\% of the Crab Nebula flux in 50 hours
\citep{aliu09}.
The observations were carried out in wobble mode,
pointing at two different symmetric regions situated $0.4^{\circ}$ away from
the source position to simultaneously evaluate the background \citep{Fomin94}.

The data were analyzed using the standard MAGIC analysis and
reconstruction software, MARS \citep{Zanin13}. The recorded images were
calibrated, cleaned, and parametrized
\citep{Hillas85,Albert2008a}. The background rejection and the estimation of the gamma-direction
was performed using the Random Forest (RF) method \citep{magic:RF}. 
The energy of each event was estimated using look-up tables generated by
Monte Carlo simulations \citep{MAGIC_stereo_performance}. For mono-observations, the event direction and energy of the primary gamma ray were
also reconstructed with the RF method. 

The LIVERPOOL robotic telescope is an optical telescope that
is also located on La Palma. Its
instrument, FRODOspec, provides spectra with R $\sim$ 5500 resolution simultaneously
in the blue and red spectral ranges. The red spectrum includes the H$\alpha$ line,
which we used as the prime indicator of the Be circumstellar disk through the 
equivalent width (EW), full width at half maximum (FWHM), and centroid velocity, the latter two obtained through a single Gaussian fit to the emission profile.

\lsi\ was observed between August 2010 and September 2014. All data were
obtained with the MAGIC stereoscopic system, except for January 2012, when MAGIC-I
was inoperative. 
The source was observed during the orbital range $\phi$ = 0.5 -- 0.75  to observe the complete trend of the periodical
outburst peak of the TeV emission, with the aim of detecting a putative long-term modulation.
Contemporaneous observations with MAGIC and LIVERPOOL were performed during orbital phases 0.75 -- 1.0, which
are the phases where sporadic VHE emission had been detected and which does not seem to present yearly periodical variability of the flux level \citep{lsi_flux_states}. Since the fluxes in this orbital period are not influenced by the long-term modulation, changes in the relative optical and TeV 
fluxes are larger and easier to measure. The aim of these contemporaneous observations is to search for (anti-)correlation between the mass-loss rate of the Be star and the TeV emission. The details of the observations in an orbit-to-orbit basis are summarized in Table~\ref{tab:LSI_observations}. 

%The cyclical behavior proposed by \citep{Torres12}, predicted that the system would be back in high state around May 2010 and then it would lower or disappear about October 2012. On this multi-year campaign, the source has been observed at super-orbital phases $\phi_{super-orbit} \sim$ 0.25 (January 2011), $\phi_{super-orbit} \sim 0.55$ (February 2012), $\phi_{super-orbit} \sim$ 0.7 (February 2013), $\phi_{super-orbit} \sim$ 0.97(January 2014) and $\phi_{super-orbit} \sim$ 0.1 (September 2014), covering as much as possible the expected 1667-day super-orbital cycle. The source was observed, every time possible, from on the orbital range $\phi$ = 0.5 -- 0.75 for three hours per night, to see the complete trend and be able to detect the outburst peak, even if the source is in a low-state of emission. For phases 0.75 -- 1.0, the source was observed two hours per night. 

%-----------------  Observations Table -------------------------

\begin{table}[t!] \scriptsize
\begin{center}
 \caption{Observations of \lsi\ performed by MAGIC. The first column denotes the orbit number, the second column indicates the dates of the observation, the third and fourth column display the orbital and super-orbital phases, while the fifth and sixth columns indicate the integration time and its distribution, respectively. \label{tab:LSI_observations}}
 \begin{tabular}{cccccc}
\hline
\hline
Orbit & MJD & $\phi_{orbital}$ & $\phi_{super-orbital}$ & Time & Number\\
number & range & range &  & hours &of days\\
\hline
\hline
1 & 55415.2 & 0.75 & 0.23 & 1.14 & 1\\
2 & 55441.2 - 55444.2 & 0.73 - 0.84 & 0.25 & 3.98 & 3\\
3 & 55471.1 & 0.86 & 0.27 & 0.76 & 1\\
4 & 55486.1 - 55500.1 & 0.42 - 0.95 & 0.28 & 3.63 & 4\\
5 & 55512.0 & 0.40 & 0.29 & 1.92 & 1\\
6 & 55543.0 & 0.57 & 0.30 & 2.06 & 1\\
7 & 55568.9 - 55574.0 & 0.55 - 0.74 & 0.32 & 10.81 & 6\\
21 & 55944.0 - 55945.0 & 0.70 - 0.74 & 0.55 & 2.56 & 6\\
22 & 55969.8 - 55977.8 & 0.68 - 0.99 & 0.56 & 3.91 & 6\\
32 & 56242.0 - 56243.0 & 0.95 - 0.99 & 0.72 & 2.20 & 2\\
33 & 56266.9 - 56267.9 & 0.89 - 0.93 & 0.74 & 2.10 & 2\\
34 & 56295.9 - 56296.8 & 0.99 - 0.01 & 0.77 & 4.04 & 2\\
44 & 56549.1 - 56550.1 & 0.54 - 0.58 & 0.91 & 5.67 & 2\\
45 & 56576.1 - 56579.1 & 0.56 - 0.67 & 0.92 & 7.90 & 4\\
46 & 56602.0 - 56607.1 & 0.54 - 0.73 & 0.94 & 9.90 & 5\\
48 & 56656.9 - 56663.9 & 0.61 - 0.87 & 0.98 & 15.65 & 8\\
%43 & 56656.9 - 56663.9 & 0.61 - 0.87 & 0.98 & 15.65 & 8\\
57 & 56900.1 & 0.79 & 0.12 & 2.22 & 1\\
58 & 56920.1 - 56930.1 & 0.54 - 0.92 & 0.13 & 20.72 & 10\\
\hline
\hline
 \end{tabular}
 \end{center}
\end{table}

\section{Results}

\subsection{Spectral stability}

{\lsi\ has shown variability on timescales of years in the strength of its 
periodic outburst peaks. 
To understand if the mechanism producing gamma rays is the same, independent of the flux of the source and its super-orbital state, it is interesting to search for spectral variability as a function of time. 
The VHE spectrum for the complete observed data set (see Table~\ref{tab:LSI_daily_MAGIC}) can be described as 
%\begin{equation}
%{{dN_{\gamma} \over {dA dt dE}}} = (4.4 \pm 0.1_{stat} \pm 0.2_{sys}) \times 10^{-13} E^{(-2.4 \pm 0.2_{stat} \pm 0.2_{sys})} TeV^{-1} cm^{-2} s^{-1}
%\end{equation}
\begin{equation}
{{\mathrm{d}N_{\gamma} \over {\mathrm{d}A \mathrm{d}t \mathrm{d}E}}} = N_{0} \left(\frac{E}{E_{0}}\right)^{\alpha}
,\end{equation}
with $N_{0} = (4.4 \pm 0.1_{stat} \pm 0.2_{sys}) \times 10^{-13} \mathrm{TeV}^{-1} \mathrm{cm}^{-2} \mathrm{s}^{-1}$, $\alpha = -2.4 \pm 0.2_{stat} \pm 0.2_{sys}$ and the normalization energy {E$_0$ = 1 {TeV}}.

%--------------------Nightly Cycle VI Lightcurve Table----------------------------
\begin{table}[t!] \scriptsize
\begin{center}
 \caption{Daily integrated flux for energies above 300 GeV of \lsi\ measured by MAGIC from 2010 to 2014. The nights with an
asterisk are those where simultaneous optical data taken by LIVERPOOL is available. Horizontal lines indicate different orbits. \label{tab:LSI_daily_MAGIC}}
 \begin{tabular}{ccccc}
 \hline
\hline
MJD & $\phi$ & Significance & Integral flux  & Time$_{eff}$\\
& & (Li$\&$Ma) & (E > 300 GeV) &\\
$[Days]$ & & [$\sigma$] & [$10^{-12} cm^{-2} s^{-1}$] & [hours]\\
\hline
\hline
55415.2* & 0.75 & 1.5 & 3.3 $\pm$ 1.6 & 1.14\\
\hline
55441.2 & 0.73 & 6.4 & 5.8 $\pm$ 1.3 & 1.94\\
55442.1* & 0.76 & 1.8 & 4.9 $\pm$ 1.7 & 1.19\\
55444.2* & 0.84 & -0.7 & -1.2 $\pm$ 0.9 & 0.85\\
\hline
55471.1* & 0.86 & 0.6 & -0.6 $\pm$ 1.2 & 0.76\\
\hline
55486.1 & 0.42 & 3.4 & 3.9 $\pm$ 1.4 & 1.12\\
55498.1* & 0.87 & 0.6 & 0.4 $\pm$ 0.7 & 1.24\\
55499.1 & 0.91 & -0.3 & -0.5 $\pm$ 1.4 & 0.38\\
55500.1 & 0.95 & 0.9 & 0.2 $\pm$ 1.4 & 0.89\\
\hline
55512.0 & 0.40 & 1.6 & 1.5 $\pm$ 1.2 & 1.92\\
\hline
55543.0 & 0.57 & 2.3 & 3.1 $\pm$ 1.0 & 2.06\\
\hline
55568.9 & 0.55 & 12.3 & 9.4 $\pm$ 1.5 & 1.87\\
55569.9 & 0.59 & 2.4 & 2.3 $\pm$ 1.1 & 1.61\\
55571.0 & 0.62 & 7.0 & 5.8 $\pm$ 1.1 & 2.62\\
55572.0 & 0.66 & 6.0 & 10.0 $\pm$ 2.3 & 1.11\\
55573.0 & 0.70 & 2.9 & 1.7 $\pm$ 1.2 & 1.13\\
55574.0 & 0.74 & 2.2 & 0.4 $\pm$ 0.3 & 2.47\\
\hline
55944.0 & 0.70 & 4.0 & 12.0 $\pm$ 3.2 & 1.70\\
55945.0 & 0.74 & 0.7 & 1.0 $\pm$ 4.2 & 0.86\\
\hline
55969.8 & 0.68 & -0.4 & -1.1 $\pm$ 1.1 & 0.92\\
55970.8 & 0.72 & 1.7 & 2.5 $\pm$ 1.9 & 0.35\\
55975.8 & 0.91 & 1.3 & 1.3 $\pm$ 0.8 & 1.66\\
55976.8 & 0.95 & 1.1 & 1.1 $\pm$ 1.9& 0.42\\
55977.8* & 0.99 & 1.3 & 1.3 $\pm$ 1.9 & 0.57\\
\hline
56242.0 & 0.95 & 0.7 &2.6 $\pm$ 1.6 & 0.99\\
56243.0 & 0.99 & 1.7 &4.5 $\pm$ 1.8 & 1.24\\
\hline
56266.9* & 0.89 & 2.3 &6.6 $\pm$ 1.9 & 1.09\\
56267.9* & 0.93 & 5.0 &7.0 $\pm$ 2.0 & 1.02\\
\hline
56295.9* & 0.99 & 3.0 &4.1 $\pm$ 1.4 & 1.71\\
56296.8* & 0.01 & -0.6 &-1.4 $\pm$ 0.9 & 2.33\\
\hline
56549.1 & 0.54 & 3.1 &6.1 $\pm$ 1.6 & 2.69\\
56550.1 & 0.58 & 0.6 &0.3 $\pm$ 1.0 & 2.98\\
\hline
56576.1 & 0.56 &  0.2 &0.8 $\pm$ 1.8 & 1.09\\
56577.0 & 0.60 & -2.7 &-1.8 $\pm$ 1.1 & 2.82\\
56578.0 & 0.63 & -2.7 &-1.8 $\pm$ 1.0 & 2.69\\
56579.1 & 0.67 &  0.2 &0.4 $\pm$ 1.5 & 1.30\\
\hline
56602.0 & 0.54 & 0.7 &1.6 $\pm$ 1.3 & 2.10\\
56603.0 & 0.58 & 2.6 &1.3 $\pm$ 1.0 & 2.93\\
56604.0 & 0.61 & 1.0 &2.3 $\pm$ 1.2 & 2.72\\
56606.0 & 0.69 & 1.6 &1.8 $\pm$ 1.6 & 0.98\\
56607.1 & 0.73 & 3.6 &5.3 $\pm$ 1.9 & 1.17\\
\hline
56656.9 & 0.61 & 0.9 & 1.2 $\pm$ 1.5 & 1.68\\
56657.9 & 0.65 & -0.1 & -0.2 $\pm$ 1.1 & 2.93\\
56658.9 & 0.69 & 2.0 & 2.7 $\pm$ 1.5 & 2.37\\
56659.9 & 0.72 & 3.0 & 4.1 $\pm$ 1.6 & 1.75\\
56660.9* & 0.76 & 1.3 & 1.7 $\pm$ 1.4 & 2.20\\
56661.9* & 0.80 & 0.1 & 0.1 $\pm$ 1.5 & 1.14\\
56662.9* & 0.84 & 0.7 & 1.0 $\pm$ 1.5 & 1.95\\
56663.9* & 0.87 & 4.6 & 5.0 $\pm$ 1.4 & 1.63\\
\hline
56900.1 & 0.78 & -0.21 & 0.9 $\pm$ 1.1 & 2.22\\
56920.1 & 0.54 & 4.92 & 4.6 $\pm$ 1.4 & 1.74\\
56921.1 & 0.58 & 9.27 & 12.1 $\pm$ 1.6 & 2.27\\
56922.1 & 0.61 & 1.43 & 2.2 $\pm$ 1.0 & 2.93\\
56923.1 & 0.65 & 6.48 & 7.2 $\pm$ 1.2 & 2.92\\
56924.1 & 0.69 & 0.88 & 0.5 $\pm$ 0.9 & 2.93\\
56925.1 & 0.73 & 0.21 & 1.7 $\pm$ 1.3 & 1.49\\
56927.1 & 0.80 & 0.12 & 1.3 $\pm$ 1.9 & 0.62\\
56928.1 & 0.84 & -0.83 & -0.5 $\pm$ 1.0 & 1.94\\
56929.1 & 0.88 & -1.49 & -0.6 $\pm$ 0.9 & 1.93\\
56930.1 & 0.92 & -1.42 & -1.8 $\pm$ 0.9 & 1.94\\
\hline
\hline
 \end{tabular}
 \end{center}
\end{table}

%The spectral index of the unfolded spectrum for the phases where the periodical TeV outburst is detected is used to determine if such variability exits. 
Table ~\ref{tab:spectral-indexes_allphases} collects the spectral indices for each of the observational campaigns. They are all similar within systematic errors. 
In 2011, a hardening was observed, but with low statistical significance. It was not confirmed by later observations
%There might have been hardening observed during 2011, but it is not statistically significant and it was not {confirmed by} later observations.
Therefore, we can conclude that the VHE spectrum is consistent with a single power-law during different epochs. }

%--------------Spectral indexes all cyclesTable-------------------------
\begin{table}[t!] \scriptsize
\begin{center}
 \caption[Spectral indices for \lsi\ campaigns]{Spectral results for the different observational campaigns of \lsi. The first column indicates the campaign name, the second the orbital range where the SED was computed, the third the average super-orbital phase of the campaign, and the fourth the spectral index with its corresponding statistical and systematic uncertainties.\label{tab:spectral-indexes_allphases}}
\begin{tabular}{cccc}
 \hline
\hline
Campaign & Orbital & Super-orbital & Spectral \\
& interval & phase &index\\
\hline
\hline
\citep{Albert_Science_LSI} & 0.4 -- 0.7 & $\sim$0.22 & $-2.6 \pm 0.2 \pm 0.2$\\
%\citep{Albert_Science_LSI} & & &\\
\citep{Anderhub09} & 0.4 -- 0.7 & $\sim$0.59 & $-2.6 \pm 0.2 \pm 0.2$\\
%\citep{Anderhub09} & & &\\
\citep{MAGIC_lsi_periodic09} & 0.6 -- 0.7 & $\sim$0.41 & $-2.6 \pm 0.2 \pm 0.2$\\
%\citep{MAGIC_lsi_periodic09} & & &\\
\citep{lsi_flux_states}  & 0.6 -- 0.8 & $\sim$0.08 & $-2.5 \pm 0.5 \pm 0.2$\\
%\citep{lsi_flux_states} & & &\\
2011 (this work) & 0.6 -- 0.8 & $\sim$0.28 & $-2.2 \pm 0.1 \pm 0.2$\\
% & & &\\
2012  (this work)& 0.7 & $\sim$0.55& $-2.7 \pm 0.5 \pm 0.2$\\
%(this work) & & &\\
2013  (this work)& 0.5 -- 0.8 &$\sim$ 0.92 &$-2.5 \pm 0.5 \pm 0.2$\\
2014  (this work)& 0.5 -- 0.8 &$\sim$ 0.13 &$-2.5 \pm 0.1 \pm 0.2$\\
%(this work)& & &\\
\hline
\hline
 \end{tabular}
 \end{center}
\end{table}

\begin{figure}[ht!]
\centering
\resizebox{\hsize}{!}{\includegraphics{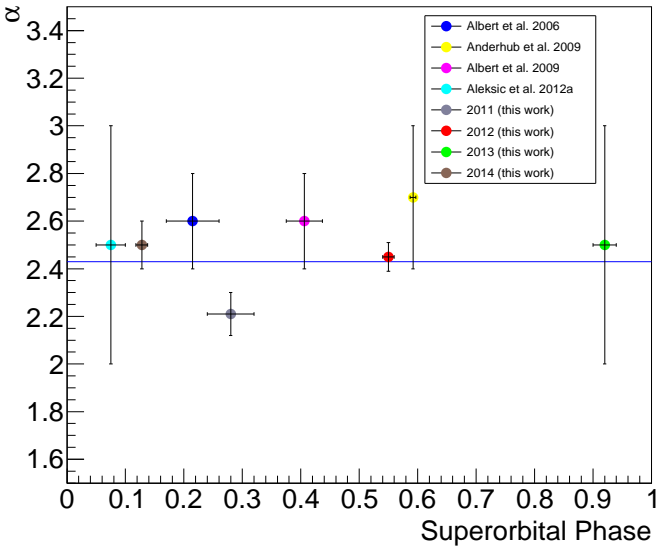}}
%\resizebox{\hsize}{!}{\includegraphics{alpha_Paper.eps}}
\caption[super-orbital dependence of the spectral index of \lsi]{Super-orbital dependence of the spectral index $\alpha$ for all MAGIC campaigns of \lsi, considering a 1667-day period. The blue line corresponds to the average value. \label{fig:alpha_superorbit_all_thesis}}
\end{figure}

The dependence of $\alpha$ on phase with the 1667 day 
super-orbital period is shown in Fig.~\ref{fig:alpha_superorbit_all_thesis}. The data can be
fit to a constant value of $2.43 \pm 0.04$ ($\chi^{2}$/dof = 8.9/7).
%Actually the systematics on the spectral index are +/-0.15, and for mono
%observations they are +/-0.2. So most of the statistical uncertainties in the
%plot are larger than the systematic uncertainties. With this good fit
%probability we can conclude that the spectral index
%compatible with being stable. \textcolor{green}{We think that when a fit is given, the
%probability of goodness shall also be given along with it, to prove the
%strength of the result obtained, which in this case it quite high.}}

We divided the dataset into intervals of high (defined as the flux being at 5
-- 10\% of the Crab Nebula flux) or low (flux at 2 -- 5\% of the Crab Nebula
flux) activity. {We also separated the data depending on the orbital interval
of the emission into a periodical ($\phi$ = 0.5 -- 0.75) interval, where the periodic TeV outburst occurs, and a {sporadic}
($\phi$ = 0.75 -- 1.0) interval, where significant emission has been detected sporadically in the past.} In neither case does the spectral index vary
significantly. Fitting the data with a constant $\alpha$ results in an average
value of $2.4 \pm 0.1$ ($\chi^{2}/dof = 0.90 / 2$) with a probability of 0.83.
%Figure~\ref{fig:alpha_superorbit_levels_thesis} shows the dependence of the spectral index with the super-orbital phase considering these divisions. The large uncertainties in the sample corresponding to 2 -- 5 $\%$ of the Crab flux and $\phi$ = 0.8 -- 1.0 is due to the low signal enclosed in these data. 
%If there exits any variation in the spectral index, it is below the sensitivity of the MAGIC telescopes.

%--------------------Spectral index vs Super-orbit Figure------------------------

\subsection{Super-orbital modulation}

{We performed a search} for indications of a super-orbital flux modulation using the peak flux measurements at orbital phases $\phi$ = 0.5 -- 0.75.
However, some of the system's emission peaks may have been missed, for instance, because observations can only be performed once every 24 hours and because {some nights were} lost because of bad weather or technical problems. To minimize this effect, 
we {only selected those} measurements for which at least three consecutive observations in the requisite phase interval were
successful and for which the middle time shows a higher flux than the others.

%{We justified this approach by simulating exactly this situation. Consider the maximum of three points of three consecutive nights at the selected orbital phases. The measurement was simulated assuming the peak to have a shape similar to the one observed in~\cite{MAGIC_lsi_periodic09}. We then calculated the percentage difference between the actual maximum of the emission and the assigned value to the maximum. With this approach we found that, at the 1$\sigma$ level, the difference between the estimated maximum flux and the actual maximum is $< 35$\%.} 

We justified this approach by simulating exactly this situation. Since observations can only be performed once every 24 hours, a maximum in emission from the binary system may be missed. Additionally, observations were also lost during intervals of bad weather or technical problems.  
To minimize the effects of missing the maximum emission during any cycle, we only considered those orbits for which observations occurred during at least three consecutive nights in the specific phase interval. We also required that the middle observation showed the highest flux. By further assuming a shape (Landau distribution) similar to the one observed in \cite{Anderhub09},  we then calculated the percentage deviation between the actual maximum of the emission and the assigned value of the maximum. We checked that at one sigma, the difference between the estimated maximum flux level and the actual maximum was smaller than $< 35$\%, while this value increased to $90$\% if the maximum was used, even requiring three consecutive nights. This estimate allowed us to study the dependence of the amplitude on the orbital periodic emission observed at VHE with time.

%the binary system may not be observed while being at its maximum of emission. Moreover, some observations may be lost due to bad weather or technical problems, increasing the probability to miss the actual peak emission. To minimize the effect of missing the maximum emission of a certain orbital, we only consider those for which at least three consecutive nights in the requested phase range were successful and the central one shows a flux larger than the others. With this approach and assuming a shape similar to the one observed in \cite{Anderhub09}, we then calculated the percentage difference between the actual maximum of the emission and the assigned value to the maximum. We ensure that, at 1 sigma, the difference between the estimated maximum flux level and the actual maximum is smaller than $< 35$\%, while if simply the maximum is taken, even requesting three consecutive nights, this value increases up to $90$\%. This estimation permits to study the dependence of the amplitude of the orbital periodic emission observed at VHE with time.}

%With this approach, and assuming a shape similar to
%the one observed in \cite{MAGIC_lsi_periodic09}, we ensure that, at the 1$\sigma$ level, the difference between the estimated maximum flux and the actual maximum is $< 35$\%. [NB: THis needs to be expanded and justified, Monte Carlo estimate?]
%{\bf check}
%while if simply the maximum is taken even requestingthree consecutive nights this difference goes up to 90\%. This allows to study the
%dependence of amplitude of the orbital periodic emission observed at VHE with time.

All archival data of \lsi\ recorded by MAGIC since its detection in 2006 \citep{Albert08_LSI,Anderhub09,MAGIC_lsi_periodic09,lsi_flux_states} 
and the data from the observing campaigns presented were folded onto the superorbital period of 1667 days. 
We considered statistical and systematic uncertainties in the integrated flux:
12\% systematics for stereo data, according to \cite{MAGIC_stereo_performance}, and 15\% systematics for mono data, estimated from \cite{MAGIC_crab_mono_2008}. We augmented our sample using the many observations {made} by VERITAS of \lsi, that is, \citet{VERITAS08, Acciari09, Acciari11} and \citet{Aliu13}, using the same procedure to identify the peak of emission. We assumed 
20$\%$ systematic uncertainties \citep{Griffin_performance} that were added quadratically to the corresponding statistical uncertainties.

\begin{figure}[t!]
%\resizebox{\hsize}{!}{\includegraphics{Phaseogram_big_labelok.eps}}
\resizebox{\hsize}{!}{\includegraphics{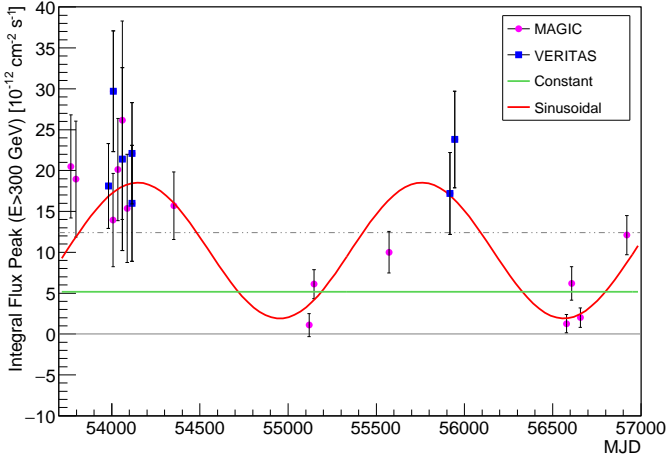}}
%\resizebox{\hsize}{!}{\includegraphics{../SuperOrbit/MAGIC-VERITAS_MJD_new.eps}}
\caption[MJD evolution]{Temporal evolution of the peak of the VHE emission for each orbital period in terms of the MJD. MAGIC (magenta dots) and VERITAS (blue squares) data within orbital phases 0.5 -- 0.75 have been considered. The gray dashed line represents 10\% of the Crab Nebula flux, the gray solid line the zero level. The fit with a sinusoidal is plotted in red and the fit with a constant in blue.}\label{fig:MAGIC-VERITAS_MJD-Paper}
%, considering an orbital period of 60.37 days and $T_{0}$ = MJD 53245.3 (from \citep{Williams10}).\label{fig:phases}}
\end{figure}

\begin{figure}[t!]
%\resizebox{\hsize}{!}{\includegraphics{Phaseogram_big_labelok.eps}}
%\resizebox{\hsize}{!}{\includegraphics{../SuperOrbit/MAGIC-VERITAS_nostat.eps}}
%\resizebox{\hsize}{!}{\includegraphics{../SuperOrbit/MAGIC-VERITAS_new.eps}}
\resizebox{\hsize}{!}{\includegraphics{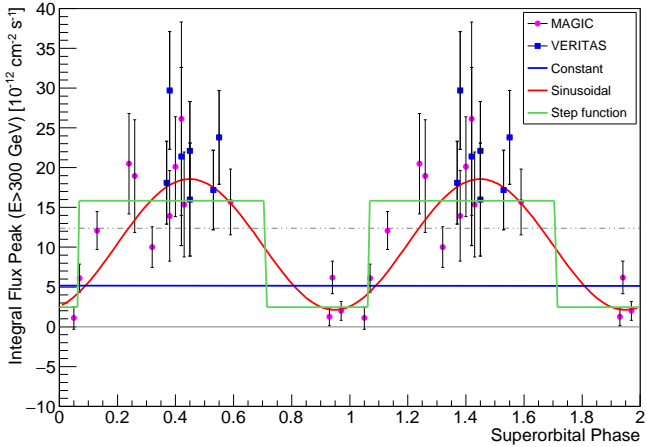}}
%\resizebox{1.03\hsize}{!}{\includegraphics{MAGIC-VERITAS-STEPF_Paper_Final.eps}}
%\resizebox{\hsize}{!}{\includegraphics{../SuperOrbit/MAGIC-VERITAS_Stepf2.eps}}
\caption[Super-orbital phase evolution]{Peak of the VHE emission in terms of the super-orbital phase defined in radio. Each data point represents the peak flux emitted in one orbital period during orbital phases 0.5 -- 0.75 and is folded into the super-orbit of 1667 days known from radio observations \citep{Gregory02}. MAGIC (magenta dots) and VERITAS (blue squares) points have been used in this analysis. The fit with a sinusoidal (solid red line), with a step function (solid green line), and with a constant (solid blue line) are also represented. The gray dashed line represents $10\%$ of the Crab Nebula flux, the gray solid line the zero level for reference.\label{fig:superorbit_peak_05075_sinlin_sys_correct_MAGIC-VERITAS}}
%, considering an orbital period of 60.37 days and $T_{0}$ = MJD 53245.3 (from \citep{Williams10}).\label{fig:phases}}
\end{figure}

The temporal evolution of the peak-integrated flux above 300\,GeV for each
orbit is presented in Fig.~\ref{fig:MAGIC-VERITAS_MJD-Paper}. Folded light-curve data were fit by first assuming a hypothesis of a constant flux, and secondly assuming a sinusoidal function.
%The first fit was done with a constant and the second one with a sine function leaving all parameters free. {red}{[NB: actually, the fit is done *relative* to a constant (you were citing statistics and probabilities before, the null hypothesis is always a constant or a constant plus noise at some level.]} 
The fit to a sine function yielded a period of $1610~\pm~58$ days with a probability of 6\% {($\chi^{2}$/dof = 26.6/17)}.
Since the fit probability for a constant fit is much lower, with $4.5 \times 10^{-12}$ {($\chi^{2}$/dof = 114.8/20)}, we can conclude that the TeV flux is variable along the superorbit and that a sinusoidal
behavior with a period of 1610 days is preferred over a constant one.
%{DH: No entiendo que quiere que hagamos aqui. Esta mal decir que haciamos el fit con un sine?} \textcolor{green}{Yo tampoco entiendo que quiere que hagamos aqui... Yo entiendo que es correcto decir que hacemos un fit a un valor constante y a un seno, y que nuestra null hypothesis sea el fit a una constante no hace que esto sea incorrecto}
%It is when compared to the fit to a constant flux (which results in probability of
%$4.5 \times 10^{-12}$ for being correct). Therefore, the flux variability being a random fluctuation can be discarded.

Figure~\ref{fig:superorbit_peak_05075_sinlin_sys_correct_MAGIC-VERITAS} shows the same data, but folded onto the super-orbital period of 1667 days. Here the data were fit not only with a constant and a sinusoidal, {but also to two-emission levels (step function):

\[
  f(x) = \left\{\begin{array}{ccl}
      a & \mathrm{if} &  p_1 < x < p_2, \\[\jot]
      b & \mathrm{if} &  x < p_1 \, \& \, x > p_2,
    \end{array}\right.
\]

{where $a$, $b$, $p_1$ , and $p_2$ are the free parameters of the fit.}
The results of the {different fits are presented} in Table~\ref{tab:LSI_results}. The probability for a constant flux is negligible, 4.5 $\times 10^{-12}$. Assuming a sinusoidal signal, the fit probability reaches 8\% ($\chi^{2}$/dof = 27.2/18). 
The fit to a step function resulted in a fit probability of 7\% ($\chi^{2}$/dof = 26.4/17).
%The data have furthermore {been fit to two-emission levels (step function)}, resulting in a fit probability of 7\% ($\chi^{2}$/dof = 26.4/17).
We furthermore quantified the probability that the improvement found when fitting a sinusoid or a step function instead of a constant is produced by chance. To obtain this probability, we considered the likelihood ratio test \citep{Mattox1996}. In both cases this chance probability is $<2.5\times10^{-10}$, which is low.
This shows that the observed intensity distribution can be described by a high and a low state and with a smoother transition. We conclude that there is a super-orbital signature in the TeV emission of \lsi\ and that it is compatible with the 4.5-year radio modulation seen in other frequencies.
%The probability for a constant flux is with $4.5 \times 10^{-12}$ negligible. 
%Having ruled out a constant emission level and assuming a sinusoidal signal, the probability reaches 8\%. 
%\textcolor{red}{[NB: just give the $\chi^2/dof$ and that's enough!]} {Alicia, tienes los valores $\chi^2/dof$ tu?} The data were also fit with a step function, resulting in a fit probability of $4.7 \times 10^{-2}$. Thus, the time series 
%cannot be described by a strictly two-state variability.  A continuous function in time is required. We conclude that there is a super-orbital signature in the TeV emission of \lsi\, and that it is consistent with the $\sim$4.5-year radio modulation.

%--------------------Results of super-orbital fit Table----------------------------
\begin{table}[t!] \scriptsize
 \caption{Fitting functions with the corresponding fit probabilities for MAGIC + VERITAS data of \lsi\ {folded into the superorbit (Fig.~\ref{fig:superorbit_peak_05075_sinlin_sys_correct_MAGIC-VERITAS}).} \label{tab:LSI_results}}
\begin{center}
\resizebox{0.3\textwidth}{!}{
 \begin{tabular}{ccc}
 \hline
\hline
Function & Fit probability & $\chi^{2}$/dof \\
\hline
\hline
Constant & $4.5 \times 10^{-12}$ & 114.8/20\\
Step function & 0.07 & {26.4/17}\\
Sinusoidal & 0.08 & 27.2/18\\
\hline
\hline
 \end{tabular}
 }
 \end{center}
\end{table}

%===============================================================================================
%===============================================================================================

\subsection{Simultaneous optical-TeV observations}

\begin{table}[t!] 
\begin{center}
 \caption[Correlation between TeV flux and $H\alpha$ parameters, phases 0.75 - 1.0]{Correlations between the TeV flux obtained by MAGIC and the $H\alpha$ parameters (EW, FWHM, and vel) measured by LIVERPOOL for the extended orbital interval 0.75 -- 1.0. Only TeV data with a significance higher than 1$\sigma$ have been considered. The first column indicates the level of simultaneity of the observations, the second column shows the parameters we used to search for a correlation, and the third and fourth columns give
 the Pearson correlation coefficient and the associated probability for a non-correlation. \label{tab:corr-prob}}
%\resizebox{0.9\textwidth}{!}
 \begin{tabular}{cccc}
 \hline
\hline
Simultaneity & Parameters & \textit{r} & Prob\\
\hline
\hline
Nightly & TeV - EW  & -0.51 & 0.04\\ %-0.28 & 0.89
Nightly & TeV - FWHM & -0.22 & 0.27\\%-0.10 & 0.67
Nightly & TeV - vel & -0.38 & 0.11\\%-0.43 & 0.97
Contemporary (hourly)   & TeV - EW  & -0.14 & 0.37\\%-0.38 & 0.86
Contemporary (hourly)   & TeV - FWHM & -0.44 & 0.16\\%-0.18 & 0.69
Contemporary (hourly)   & TeV - vel & -0.21 & 0.35\\%-0.43 & 0.89
%Nightly        & TeV - EW  & -0.23 & 0.84\\ %-0.28 & 0.89
%Nightly        & TeV - FWHM & -0.14 & 0.72\\%-0.10 & 0.67
%Nightly        & TeV - vel & -0.44 & 0.97\\%-0.43 & 0.97
%Contemporary (hourly)  & TeV - EW  & -0.32 & 0.80\\%-0.38 & 0.86
%Contemporary (hourly)  & TeV - FWHM & -0.24 & 0.74\\%-0.18 & 0.69
%Contemporary (hourly)  & TeV - vel & -0.45 & 0.90\\%-0.43 & 0.89
%Strict         & TeV - EW  & -0.25 & 0.58\\%-0.32 & 0.44
%Strict         & TeV - FWHM & 0.40 & 0.53\\%0.51 & 0.24
%Strict         & TeV - vel & 0.95 & 0.24\\%0.84 & 0.19
\hline
\hline
 \end{tabular}
 \end{center}
\end{table}

The correlation between the TeV flux measured by MAGIC and the H$\alpha$ parameters measured with the  telescope (EW, FWHM, profile centroid velocity) {were determined} including statistical and systematic uncertainties and the weighted \textit{\textup{Pearson correlation coefficient}} \citep{Pearsoncoeff1917}. 

{To search for correlations, we classified the data into three different categories: 
\begin{itemize}
\item \textit{simultaneous}: The optical observations were performed
precisely during the period when MAGIC was performing its observations. Only three such points were obtained under this condition.
Because of the scarcity of such data we do not provide any correlation
coefficient for this group. 
\item \textit{contemporary (hourly)}: Data have a time difference of\textup{\textup{ {\it \textup{three hours at most}}}}.
\item \textit{nightly}: Data were obtained during the same night.
\end{itemize}
}

Only TeV data points with a significance higher than 1~$\sigma$ were considered.

% To search for correlations, we distinguished between 
%three \textit{degrees} of
%simultaneity%
% [NB: WHAT? Degrees of simultaneity? Is this like "Six Degrees of Separation"? I think you mean that you treated simultaneous, contemporary (hourly), and nightly. Say it.] have been considered: (a) when the optical observations were obtained precisely during the period when MAGIC was performing its observations (denoted as \textit{strict} observations, only three such data points were obtained, and because of the scarcity of such data we do not provide any correlation coefficient for this group); (b) when data have a time difference of {\it at most 3 hours} (denoted as {\it 3-hour}), and (c) data that were obtained during the same night (denoted as \textit{nightly}). 
The data used for the correlation are given in Table~\ref{tab:corr-prob}.  
No statistically significant correlation was found for the sample at orbital phases $\phi$ = 0.75 -- 1.0. {A hint of a correlation is observed, but its significance is low. A stronger correlation might be blurred as a result of the fast variability of the optical parameters on short timescales compared to the long exposure times required by MAGIC, and as a result of the relatively large uncertainties and small number of data points used for this analysis}.
 Figure~\ref{fig:TeV_optical_parameters} shows the H$\alpha$ measurements plotted against the TeV flux}.

%===============================================================================================
%                 Figure 5
%===============================================================================================
\begin{figure}[t!]
%%%\resizebox{\hsize}{!}{\includegraphics{All_sys_ok_paper.eps}}
%%\resizebox{\hsize}{!}{\includegraphics{Correlation_plot.eps}}
%\resizebox{1.1\hsize}{!}{\includegraphics{Correlation_Paper.eps}}
\resizebox{1.1\hsize}{!}{\includegraphics{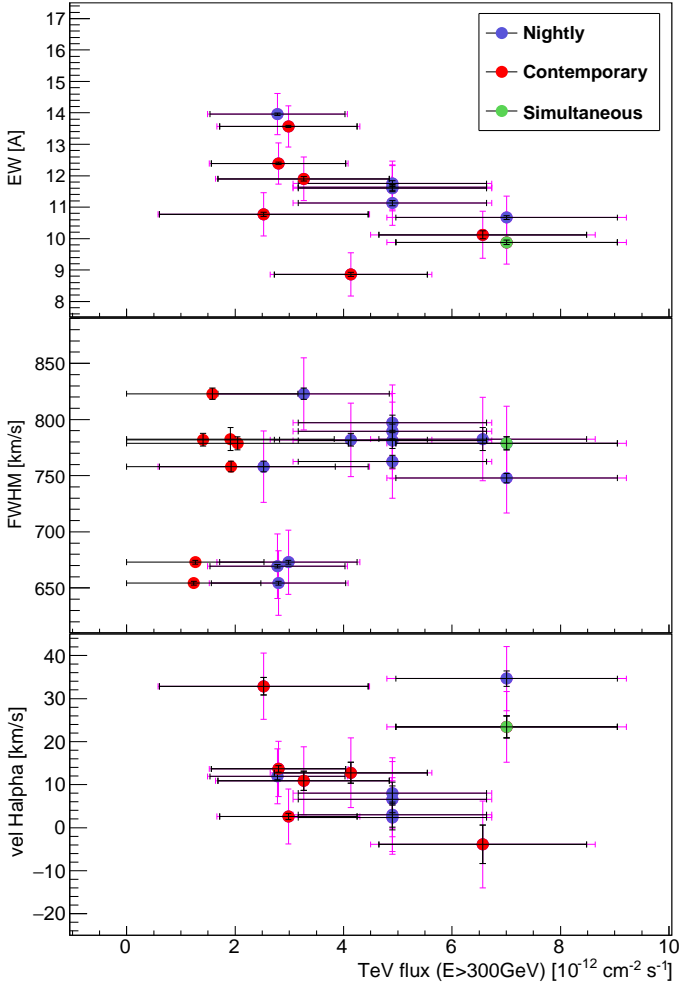}}
\caption{\label{fig:TeV_optical_parameters}Correlations between the TeV flux obtained by MAGIC and the $H\alpha$ parameters (from top to bottom: EW, FWHM, and centroid velocity (vel)) measured by the  LIVERPOOL Robotic Telescope for the orbital interval 0.75 -- 1.0. Only TeV data points with a significance higher than 1$\sigma$ have been considered. Each data point represents a ten-minute observation in the optical and a variable integration in the TeV regime: nightly (blue), contemporary (red), and strictly simultaneous data (green). Black error bars represent statistical uncertainties, while systematic uncertainties are plotted as magenta error bars.}
\end{figure}
%===============================================================================================
%===============================================================================================

\section{Discussion and conclusion}

The main conclusions from this multi-year analysis of \lsi\ TeV observations are listed below.
\begin{enumerate}

\item {We achieved a first detection of super-orbital
variability in the TeV regime. {Using} the new VHE data and the MAGIC and
VERITAS archival data, we found that the super-orbital signature of \lsi\
is consistent with the 1667-day radio period within 8$\%$. The TeV period
obtained when fitting the data to a sine function with a free frequency instead of performing the fit as before with a fixed super-orbital period known from radio, is
1610 $\pm$ 58 days with a fit probability of 6$\%$}.

%the super-orbital signature found in this work is consistent with the 1667-day radio period to within 8$\%$. The TeV period obtained when fitting the data to a sinusoidal function with free frequency is 1610 $\pm$ 58 days with a fit probability of 6$\%$
%Fitting the data with a sinusoidal function with free frequency, we obtain 
%a TeV period of 1610 $\pm$ 58 days with a fit probability of 6$\%$. \textcolor{red}{[NB: these appear to be inconsistent, reword this.]}

\item There is no statistically significant intra-day correlation between H$\alpha$ line properties and TeV emission, nor is there an obvious trend connecting the two frequencies. 

%{A hint of correlation is observed, however its low significancy makes imposible to claim any clear connection}.

\end{enumerate}

{The flip-flop model \citep{Zamanov2001,Torres12,Papitto_2012} considers \lsi\ as a pulsar-Be star binary that changes accretion states from a propeller regime during periastron to an ejector regime at apastron. The change of state is thought to be driven by the influence of matter. The higher the pressure of this matter on the (hypothesized) neutron star magnetosphere, the more likely the latter will be compressed and disrupted. When this occurs, the pulsar wind is affected or disappears, and the inter-wind shock, which is thought to contribute to the multi-frequency non-thermal emission, disappears as well.
Even when the neutron star passage at periastron is able to cut the disk off, mass accumulation will be higher in periods in which the decretion rate is higher, reaching farther out in the orbit. A larger EW of H$\alpha$ may be assumed as a proxy for such a situation. If this is the case, the optical emission {should} be anti-correlated with the TeV flux.
The detection of TeV gamma-ray super-orbital emission conforms to the predicted long-term behavior of the flip-flopping model \citep{Torres12}. The source was found in high and low states, when expected. This result extends the earlier indications for this phenomenology found using smaller samples of TeV data \citep{Li_12}.
However, we do not see this (anti-) correlation in the intra-day scales we
tested. It may be that the EW of H$\alpha$ is not the best tracer for decretion
disk size or mass accumulation, or simply that the fast and extreme changes (by
up to a factor of several in the same day) in the H$\alpha$ data of the source
and the vastly different integration times (minutes as
compared with several hours) needed to claim a detection in both
frequencies prevent us from measuring any possible trend.

\begin{acknowledgements}
We would like to thank
the Instituto de Astrof\'{\i}sica de Canarias
for the excellent working conditions
at the Observatorio del Roque de los Muchachos in La Palma.
The support of the German BMBF and MPG,
the Italian INFN, 
the Swiss National Fund SNF,
and the ERDF funds under the Spanish MINECO
is gratefully acknowledged.
This work was also supported
by the CPAN CSD2007-00042 and MultiDark CSD2009-00064 projects of the Spanish Consolider-Ingenio 2010 programme,
by grant 127740 of the Academy of Finland,
by the Croatian Science Foundation (HrZZ) Project 09/176,
by the DFG Collaborative Research Centers SFB823/C4 and SFB876/C3,
and by the Polish MNiSzW grant 745/N-HESS-MAGIC/2010/0.
J.C., and D.F.T. acknowledge support by the Spanish Ministerio de Econom\'ia y Competividad (MINECO) under grant AYA2010-18080, 
and by MINECO and the Generalitat de Catalunya under grants AYA2012-39303 and SGR 2014-1073, respectively. 

\end{acknowledgements}

%%%%%%%%%%%%%%%%%%%%%%%%%%%%%%%%%%%%%%%%%
%% bibliography
%%%%%%%%%%%%%%%%%%%%%%%%%%%%%%%%%%%%%%%%%
%\begin{thebibliography}{}
%\end{thebibliography}

\bibliography{LSI_internal_1}
\bibliographystyle{aa}

\end{document}